\documentclass[aps,twocolumn,superscriptaddress]{revtex4}
\usepackage{graphicx}

\newcommand{\ket}[1] {\left| #1 \right\rangle}
\newcommand{\prodket[2]} {\left| #1 \right\rangle \otimes
\left| #2 \right\rangle}

\begin{document}

\title{Bell's experiment with intra- and inter-pair entanglement:
Single-particle mode entanglement as a case study}

\

\author{S. Ashhab}
\affiliation{Advanced Science Institute, The Institute of Physical
and Chemical Research (RIKEN), Wako-shi, Saitama 351-0198, Japan}
\affiliation{Physics Department, Michigan Center for Theoretical
Physics, The University of Michigan, Ann Arbor, Michigan
48109-1040, USA}

\author{Koji Maruyama}
\affiliation{Advanced Science Institute, The Institute of Physical
and Chemical Research (RIKEN), Wako-shi, Saitama 351-0198, Japan}

\author{\v Caslav Brukner}
\affiliation{Fakult\"at f\"ur Physik, Universit\"at Wien,
Boltzmanngasse 5, 1090 Wien, Austria}
\affiliation{Institut f\"ur
Quantenoptik und Quanteninformation (IQOQI), \"Osterreichische
Akademie der Wissenschaften, Boltzmanngasse 3, 1090 Wien, Austria}

\author{Franco Nori}
\affiliation{Advanced Science Institute, The Institute of Physical
and Chemical Research (RIKEN), Wako-shi, Saitama 351-0198, Japan}
\affiliation{Physics Department, Michigan Center for Theoretical
Physics, The University of Michigan, Ann Arbor, Michigan
48109-1040, USA}

\date{\today}

\begin{abstract}
Theoretical considerations of Bell-inequality experiments usually
assume identically prepared and independent pairs of particles.
Here we consider pairs that exhibit both intra- and inter-pair
entanglement. The pairs are taken from a large many-body system
where all the pairs are generally entangled with each other. Using
an explicit example based on single mode entanglement and an
ancillary Bose-Einstein condensate, we show that the
Bell-inequality violation in such systems can display statistical
properties that are remarkably different from those obtained using
identically prepared, independent pairs. In particular, one can
have probabilistic violation of Bell's inequalities in which a
finite fraction of all the runs result in violation, even though
there could be no violation when averaging over all the runs.
Whether or not a particular run of results will end up being local
realistically explainable is ``decided" by a sequence of quantum
(random) outcomes.
\end{abstract}

% PACS No.s:
% 03.65.Ud: Entanglement and quantum nonlocality (e.g. EPR paradox,
%           Bell's inequalities, GHZ states, etc.)
% 03.67.Bg: Entanglement production and manipulation
% 03.75.Gg: Entanglement and decoherence in Bose–Einstein condensates

\maketitle

\section{Introduction}

Entanglement has for decades attracted interest and caused
controversy in the physics literature \cite{EntanglementReviews}.
Studies on this subject usually involve discussions of Bell's
inequality (or inequalities) \cite{Bell}, where measurement
results cannot be described by a local realistic model. One
typically thinks of a Bell-test experiment in terms of a large
number of identical pairs, each of which is composed of two
particles. One can usually understand such an experiment by
analyzing the behaviour of a single pair. The single-pair analysis
is then cast in statistical terms in order to deduce the outcome
of the whole experiment, as is usually done in quantum mechanics
(By `whole experiment' we mean an experiment that involves
repeating the single-pair procedure a large number of times, thus
obtaining statistical information about the measurement outcomes).

Here we are interested in the situation where the pairs that are
measured for the Bell test are non-identical and furthermore
entangled with each other. We start by pointing out that although
most studies assume identically prepared and independent pairs in
the Bell test, no such requirement is used or needed in the
derivations of Bell's inequalities. We then analyze the intra-pair
quantum correlations that can be observed in the presence of
inter-pair quantum correlations. This analysis is somewhat related
to, but clearly distinct from, recent studies that generalize
Bell's inequality in order to probe multipartite entanglement
\cite{Mermin,Laloe}. Using a specific physical entangled state as
an example, we show that rich statistical properties can be
obtained in the presence of inter-pair entanglement, in particular
a probabilistic violation of Bell's inequalities (of the bipartite
form). We discuss whether this probabilistic violation can be used
to exclude explanations based on local-hidden-variable (LHV)
models.

\section{The assumption of identical, independent pairs in a Bell test}

We start by noting the point that most studies of entanglement
assume the preparation of identical, independent pairs, whereas in
practice the pairs are typically generated by the same source and
can in principle be correlated. The question is therefore whether
the preparation of the pairs by a single source constitutes any
`loophole' in interpreting the violation of Bell's inequalities as
evidence against local realism. Inspection of the derivation of
Bell's inequalities \cite{Clauser}, however, shows that no
particular assumption is made on the possibility of correlations
between the different pairs. Inter-pair correlations or
entanglement therefore do not constitute any conceptual hurdle to
the interpretation of a Bell-inequality violation as evidence
against local realism.

A subtler point, which will be illustrated by an example below,
arises in the case when the ensemble average over a large number
of experiments does not violate Bell's inequality but a finite
fraction of the experimental runs (each one of which involves a
large number of pairs) do violate the inequality. As long as the
number of pairs in a given run is large enough to render the
expected statistical fluctuations around the mean negligible, such
a probabilistic violation can be used as evidence to exclude LHV
models. The crucial point here is that within the single
experimental runs that are accepted, no pairs are excluded
(e.g.~based on the measurement settings or the outcome of their
measurements as in the detection loophole). One must also keep in
mind that here we are talking about a finite fraction of all the
runs resulting in violation, whereas statistical fluctuations can
explain a violation in a small fraction of the runs, and the size
of this fraction decreases and approaches zero for large numbers
of pairs in each run. In short, LHV models would predict that no
violation of Bell's inequality can be observed in any experimental
run (in the limit that the number of pairs in one run approaches
infinity), and therefore a violation in only a finite fraction of
the runs can be used to exclude LHV models. One can also say that
typical sub-ensemble-selection loopholes in Bell-inequality tests
involve the assumption that the rejected runs would have produced
the same correlations as the accepted runs if they had been
recorded properly. In contrast, here we are incorporating all the
runs (both the accepted and rejected ones) into our analysis,
somewhat similarly to what was done in Ref.~\cite{Popescu}.

\section{Statistics of Bell-test results}

\begin{figure}[h]
\includegraphics[width=5.0cm]{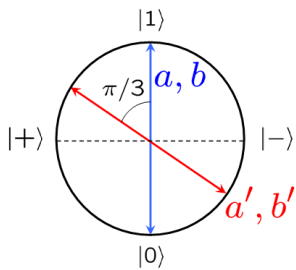}
\caption{(Color online) The measurement bases used in the Bell
test. The states $\ket{\pm}$ are defined as $\ket{\pm}=(\ket{0}
\pm \ket{1})/\sqrt{2}$. The ideal entangled state is
$(\ket{10}+\ket{01})/\sqrt{2}$. The measurement bases for the
first mode (represented by the first symbol inside the ket) are
denoted by $a$ and $a'$, and the measurement bases for the second
mode (represented by the second symbol inside the ket) are denoted
by $b$ and $b'$. Note that the measurement bases shown above,
which we choose in order to simplify our numerical calculations
below, differ from the ones that produce maximal violation of
Bell's inequality \cite{JustifyBases}.}
\end{figure}

We now briefly discuss the statistical properties of the Bell-test
violation. We use the version of the Bell test where the ideal
pair state is the state
\begin{equation}
\ket{\Phi} = \frac{1}{\sqrt{2}} \left(\ket{10}+\ket{01}\right),
\label{eq:Psi_after_BS_Fields}
\end{equation}
and the measurement bases are taken to be
$\left\{\ket{0},\ket{1}\right\}$ (to which we refer as $a$ and
$b$, depending on the sub-system on which the measurement is
performed) and $\left\{
\cos\frac{\pi}{3}\ket{0}+\sin\frac{\pi}{3}\ket{1},
\sin\frac{\pi}{3}\ket{0}-\cos\frac{\pi}{3}\ket{1} \right\}$ (to
which we refer as $a'$ and $b'$) \cite{Clauser}. This choice of
bases, which is illustrated in Fig.~1, simplifies the analysis
below \cite{JustifyBases}. A large number $M$ of pairs are
generated and measured in randomly chosen pairs of bases.
Correlation functions given by the statistical averages
\begin{equation}
C_{\alpha,\beta} = \langle \sigma_{\alpha} \sigma_{\beta} \rangle
\end{equation}
are recorded, where $\sigma_{\alpha}$ and $\sigma_{\beta}$ are the
Pauli operators along the directions $\alpha$ and $\beta$ (note
that the first and second operators affect the first and second
modes, respectively; note also that symbols such as
$\sigma_\alpha$ in our notation are sometimes expressed as
$\vec{\alpha}\cdot\vec{\sigma}$ etc. in the literature). One then
evaluates the quantity \cite{Clauser}
\begin{equation}
S = \left| - C_{a,b} + C_{a,b'} + C_{a',b} + C_{a',b'} \right|.
\end{equation}
Bell's inequality (in the Clauser-Horne-Shimony-Holt form) is now
expressed as
\begin{equation}
S \leq 2.
\end{equation}
The state in Eq.~(\ref{eq:Psi_after_BS_Fields}) gives the
expectation value $S=5/2$, thus violating the inequality
\cite{JustifyBases}. Because of the statistical nature of $S$, one
expects to obtain a different value every time the experiment is
performed (with each single experimental run involving a number
$M$ of pairs). However, if the number of pairs $M$ in a single run
is sufficiently large and the pairs are identical and
uncorrelated, the experimentally obtained value of $S$ will, with
high probability, be very close to the theoretically calculated
single-pair expectation value, with statistical variations of the
order of $1/\sqrt{M}$. In this case, $S$ becomes essentially
predictable deterministically. In the situation that we consider
in this paper, however, all the pairs are entangled with each
other. As a result, we are interested in the statistics of the $S$
values that one can expect to obtain in individual experimental
runs.

\section{Specific system: Single-particle entanglement}

\begin{figure}[h]
\includegraphics[width=8.0cm]{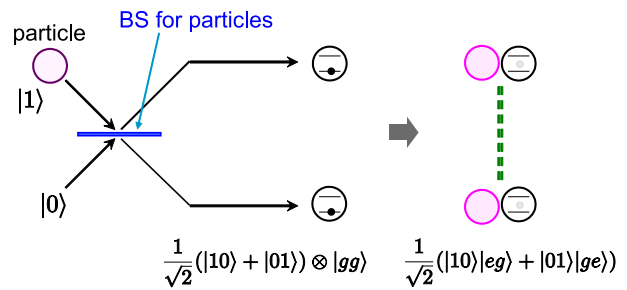}
\caption{(Color online) A schematic diagram of the setup under
consideration. After `passing through a beam splitter' (BS), a
flying particle is in a quantum superposition of being in one of
two outgoing paths. At each of the two possible final destinations
of the flying particle there is a target particle that is
initially in its ground state $\ket{g}$ and is only driven to its
excited state $\ket{e}$ if the flying particle arrives at its
location.}
\end{figure}

From now on we focus on a specific physical system as a
demonstrative example of the interesting statistics that can be
obtained using a multipartite entangled state. The system that we
consider possesses so-called mode entanglement
\cite{Tan,Ashhab1,Ashhab2,Hessmo,Devitt}. Our starting point is
the single-particle state given in
Eq.~(\ref{eq:Psi_after_BS_Fields}) where the first and second
quantum numbers now represent the number of particles in two modes
that are localized at two spatially separated, and ideally
distant, locations (see Fig.~2). We refer to such delocalized
particles as flying particles, since we imagine these particles
being emitted from a common source somewhere between the two
measurement locations. For the purposes of the present analysis,
we consider the case where these flying particles cannot be
created or annihilated. In other words, there is a conservation
law constraining the total number of particles of the
flying-particle species to be fixed. As discussed in
Refs.~\cite{Ashhab1,Ashhab2}, we imagine that a flying particle
can excite a two-level target particle from its ground state to
its excited state. By placing one target particle on each side of
the proposed setup, one can prepare the state
\begin{equation}
\ket{\Psi} = \frac{1}{\sqrt{2}} \left( \prodket[10]{eg} +
\prodket[01]{ge} \right).
\label{eq:Psi_after_excitation}
\end{equation}
One can wonder whether the preparation of the state in
Eq.~(\ref{eq:Psi_after_excitation}) poses any difficulties related
to the non-conservation of energy. In principle, energy
conservation is not a fundamental difficulty here: one can imagine
that the presence of the flying atom modifies the energy levels of
the target atom, such that an applied laser field is resonant with
the target atom only when the flying atom is on the same side of
the apparatus. One can also imagine alternative scenarios where
there is no energy difference between the states $\ket{g}$ and
$\ket{e}$.

If one performs a measurement on the target particles, the outcome
will be consistent with a reduced density matrix where the
flying-particle degrees of freedom are traced out. In the basis
$\left\{ \ket{gg}, \ket{ge}, \ket{eg}, \ket{ee} \right\}$,
Eq.~(\ref{eq:Psi_after_excitation}) gives the density matrix
\begin{equation}
\rho_{\rm TP} = \frac{1}{2} \left(
\begin{array}{cccc}
0 & 0 & 0 & 0 \\
0 & 1 & 0 & 0 \\
0 & 0 & 1 & 0 \\
0 & 0 & 0 & 0 \\
\end{array}
\right).
\end{equation}
This density matrix describes a statistical mixture of the states
$\ket{ge}$ and $\ket{eg}$ with no phase coherence between them,
i.e. with no entanglement. The reason for the lack of phase
coherence is the fact that the which-path information about the
location of the excited target particle is also carried by the
flying particle.

\subsection{Bose-Einstein condensate as an ancillary phase
reference}

We now consider an additional resource in the form of a number $N$
of particles of the same species as the flying particle. These
ancillary particles are prepared in the state
\begin{equation}
\ket{\Psi_{\rm BEC}} = \sum_{j=0}^{N} \sqrt{P_j} e^{\varphi_j}
\ket{j,N-j}_{\rm anc},
\label{eq:BEC_initial_state}
\end{equation}
where the two quantum numbers represent the number of ancillary
particles in two modes, each of which is localized on one side of
the setup (the subscript BEC indicates that we are generally
assuming $N$ to be a large number, thus forming a Bose-Einstein
condensate). For the case where the two condensates form a single
BEC state with equal weights on the two sides and zero relative
phase between them \cite{Leggett}, the distribution function in
Eq.~(\ref{eq:BEC_initial_state}) is given by
\begin{equation}
P_j = \frac{1}{2^N} \times \frac{N!}{j! (N-j)!}
\end{equation}
and the phases $\varphi_j=0$.

The state of the entire system is now given by
\begin{eqnarray}
\ket{\Psi} & = & \sum_{j=0}^{N} \sqrt{P_j} \ket{j,N-j}_{\rm anc}
\nonumber \\ & & \hspace{1cm} \otimes \frac{1}{\sqrt{2}} \left(
\prodket[10]{eg} + \prodket[01]{ge} \right).
\label{eq:Initial_State_before_disposal}
\end{eqnarray}
Following the procedure proposed in Ref.~\cite{Ashhab1}, one can
manipulate the state in
Eq.~(\ref{eq:Initial_State_before_disposal}), controllably and
coherently injecting the flying particle into the condensate of
particles that are indistinguishable from the flying particle (the
application of the injection operation is conditioned on the state
of the target atom, ensuring that unitarity is not violated), and
obtain the state
\begin{eqnarray}
\ket{\Psi} & = & \sum_{j=0}^{N} \sqrt{\frac{P_j}{2}} \Large(
\ket{j+1,N-j}_{\rm anc} \otimes \ket{eg} \nonumber \\
& & \hspace{2cm} + \ket{j,N-j+1}_{\rm anc} \otimes \ket{ge}
\Large).
\label{eq:Initial_State_after_disposal}
\end{eqnarray}
If we now trace out the state of the BEC, we find that the target
particles are described by the reduced density matrix
\begin{equation}
\rho_{\rm TP} = \frac{1}{2} \left(
\begin{array}{cccc}
0 & 0 & 0 & 0 \\
0 & 1 & \gamma & 0 \\
0 & \gamma & 1 & 0 \\
0 & 0 & 0 & 0 \\
\end{array}
\right),
\label{eq:RhoTP}
\end{equation}
where for large $N$ (and, somewhat coincidentally, for $N=1$)
\begin{equation}
\gamma = 1-\frac{1}{2N}.
\end{equation}
This state is entangled and allows the violation of Bell's
inequality (the violation is obtained even for $N=1$, as was
explained in detail in Ref.~\cite{Ashhab2}). It is worth
mentioning that for density matrices of the form of
Eq.~(\ref{eq:RhoTP}) the concurrence \cite{Wootters} is equal to
$\gamma$.

\begin{figure}[h]
\includegraphics[width=8.0cm]{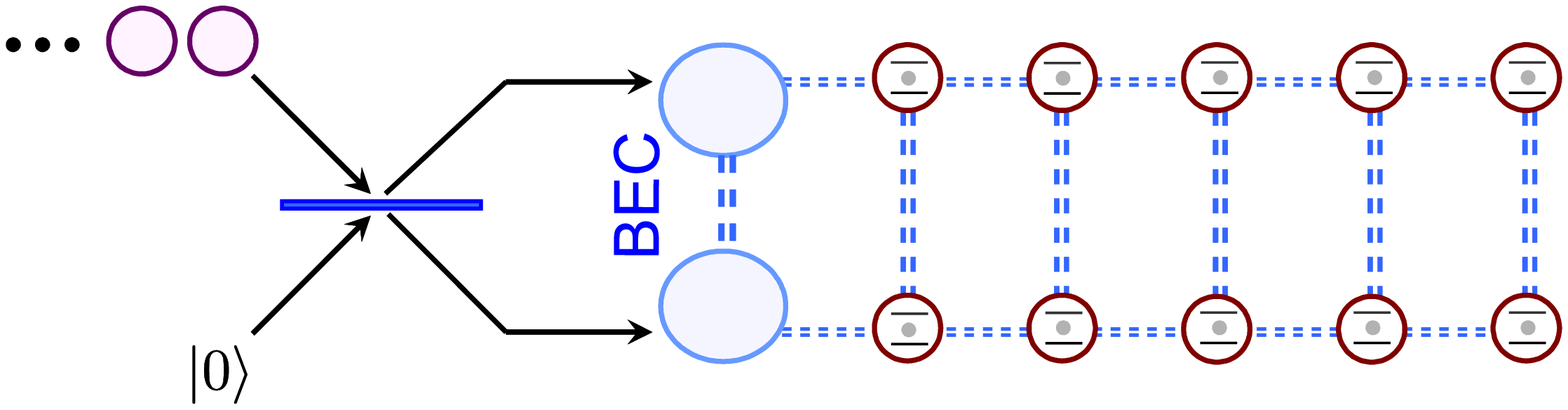}
\caption{(Color online) A schematic diagram of a single ancillary
BEC that is reused to transfer the mode entanglement of a stream
of flying particles to the internal states of the target
particles.}
\end{figure}

The above results describe the preparation of a single pair of
entangled target particles. We now consider what happens when the
same BEC is used (or rather reused) to prepare additional pairs of
entangled target particles, as illustrated in Fig.~3. Tracing out
the target-particle degrees of freedom from
Eq.~(\ref{eq:Initial_State_after_disposal}) results in a mixed
state of the BEC. As was discussed in
Refs.~\cite{Ashhab1,Ashhab2}, however, if the measurement on the
target particles is made in the $\{\ket{g},\ket{e}\}$ basis, each
one of the two possible final BEC states has the same power as the
original BEC in terms of generating entangled pairs. As a result,
any time the measurement basis $\{\ket{g},\ket{e}\}$ is used (even
if it is only for one particle in the pair), the BEC is unaffected
for the purpose of preparing more entangled pairs. Only when the
rotated basis is used for both particles does the BEC undergo
nontrivial evolution (averaging over the different outcomes
results in the same entangling power as the original BEC
\cite{EntanglementArgument}). We analyze this evolution next.

To illustrate the evolution of the ancillary resource following
the measurement of the first pair (or pairs) in the system,
instead of a BEC we take a single ancillary particle in the state
$\left( \ket{10} + \ket{01} \right)/\sqrt{2}$. Equation
(\ref{eq:Initial_State_after_disposal}) now takes the form
\begin{eqnarray}
\ket{\Psi} & = & \frac{1}{2} \Large( \ket{20}_{\rm anc} \otimes
\ket{eg} + \ket{11}_{\rm anc} \otimes \ket{eg} \nonumber
\\
& & \hspace{0.1cm} + \ket{11}_{\rm anc} \otimes \ket{ge} +
\ket{02}_{\rm anc} \otimes \ket{ge} \Large).
\end{eqnarray}
The reduced density matrix for the target-particle pair therefore
corresponds to a mixed state of the form of Eq.~(\ref{eq:RhoTP})
with $\gamma=1/2$, which means that the first pair is entangled
with concurrence equal to $1/2$. If one or both of the target
particles are measured in the $\left\{\ket{g},\ket{e}\right\}$
basis, the ancillary modes are projected onto the state $\left(
\ket{20} + \ket{11} \right)/\sqrt{2}$ or the state $\left(
\ket{11} + \ket{02} \right)/\sqrt{2}$, depending on the outcome of
the measurement. Obviously, each one of these states has the same
number distribution as the original ancillary state, apart from an
overall shift (thus these states have the same entangling power as
the original ancillary state). Let us consider, however, what
happens if both target particles are measured in the
$\left(\ket{g}\pm\ket{e}\right)/\sqrt{2}$ basis (Note that this
basis is different from the one that we use for analyzing the
Bell-test experiment but are simpler to analyze in this argument).
If the same measurement outcome is obtained for both target
particles of the first pair (a situation that occurs with
probability 3/4), the ancillary modes are projected onto the state
$\left( \ket{20} + 2 \ket{11} +\ket{02} \right)/\sqrt{6}$. This
state can then be used to prepare a second target-particle pair
with $\gamma=2/3$, i.e.~higher than the value of $\gamma$ for the
first pair. The ancillary resource is therefore enhanced when this
outcome is observed. If, on the other hand, opposite measurement
outcomes are obtained for the two target particles of the first
pair (this situation occurs with probability 1/4), the ancillary
modes are projected onto the state $\left( \ket{20} - \ket{02}
\right)/\sqrt{2}$. This state would give $\gamma=0$ for the second
pair, i.e.~the second pair would show no sign of entanglement.
Thus, although averaging over a large ensemble gives the same
value of $\gamma$ for the second pair (note here that $\frac{3}{4}
\times \frac{2}{3} + \frac{1}{4} \times 0 = \frac{1}{2}$),
knowledge of the first-pair measurement outcome gives additional
information about the entanglement in the second pair. This
phenomenon is a clear indication of correlation between the
different pairs. Naturally, this argument applies to all other
pairs that are prepared later in a long sequence as well.

\subsection{Bell test without an ancillary condensate: Probabilistic violation}

We now turn from the above argument concerning two correlated
pairs to analyzing a full Bell-test experiment involving $M$
entangled pairs. We first note that for the three choices $(a,b)$,
$(a,b')$ and $(a',b)$ [which are defined in Fig.~1], the
measurement outcomes do not depend on the value of $\gamma$ and
they give
\begin{equation}
- C_{a,b} + C_{a,b'} + C_{a',b} = 2,
\end{equation}
up to statistical fluctuations of order $1/\sqrt{M}$ that we
ignore here. The condition for the violation of Bell's inequality
therefore reduces to
\begin{equation}
C_{a',b'} > 0.
\end{equation}

\begin{figure}[h]
\includegraphics[width=7.0cm]{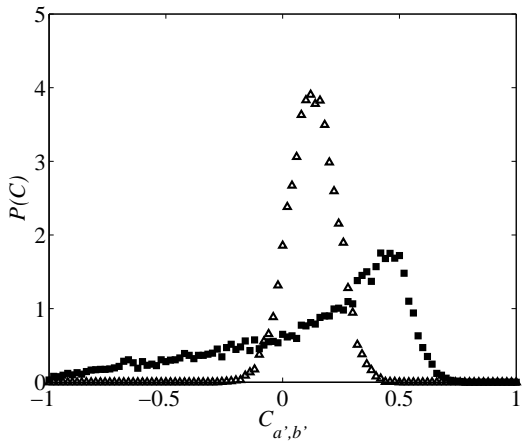}
\caption{Probability density $P(C)$ for obtaining a given value of
the correlation function $C_{a',b'}$ in a Bell test following our
procedure of reusing the ancillary source (squares). Positive
values of $C_{a',b'}$ correspond to a violation of Bell's
inequality. Here the ancillary source has $N=1$, i.e. a single
ancillary particle shared between the two sides of the setup. For
comparison we show the probability density if a new ancillary
particle was used for each entangled pair (triangles). The results
were obtained by constructing a histogram from $10^4$ runs, with
$M=400$ entangled pairs generated in each run (In the calculation
we have assumed that exactly one fourth of these pairs are
measured in the $a'$-$b'$ bases). Doubling the value of $M$
reduces the width of the curve that corresponds to the case of
independent pairs (triangles) by a factor of $\sqrt{2}$ but leaves
the curve that corresponds to the case of correlated pairs
(squares) essentially unchanged.}
\end{figure}

We now consider a single ancillary particle, and we take a stream
of flying particles used to produce a large number of entangled
target-particle pairs (note that the same ancillary particle,
along with the flying particles that are injected into the same
modes, are reused to prepare all the entangled pairs). The
probability distribution for the values of the correlation
function $C_{a',b'}$ that would be observed in experiment is shown
in Fig.~4. Unlike the prediction (depicted by the triangles) for
identical, independent pairs described by Eq.~(\ref{eq:RhoTP})
with $\gamma=1/2$, the distribution is broad, and the width
reaches a constant value for large $M$ (we have verified this
statement by comparing the results for $M=200$, $400$ and $800$).
The average value of $C_{a',b'}$ is the same for both cases, as
expected from the fact that, on average, the ancillary resource is
neither enhanced nor destroyed after repeated use
\cite{Ashhab1,Ashhab2}.

\begin{figure}[h]
\includegraphics[width=7.0cm]{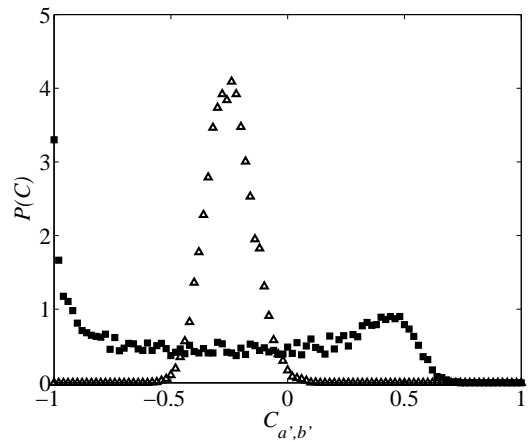}
\caption{Same as in Fig.~4, but with $N=0$, i.e.~no ancillary
particles are used. Note that the finite violation probability in
the case of independent pairs (triangles) is a result of the fact
that we use a finite number of particles in the numerical
calculations. This probability decreases and approaches zero if
the number of particles is increased. We have verified, however,
that the shape of the curve defined by the squares is essentially
unchanged if we change the number of particles (provided that this
number is much larger than one).}
\end{figure}

We next consider the case where no ancillary particles are used
($N=0$; see Fig.~5). This case is perhaps the most relevant one to
the probabilistic violation of Bell's inequality, which is the
main topic of this paper. In this case one would observe a
violation in approximately 40\% of the runs and no violation in
approximately 60\% of the runs, with the average over all the runs
being on the non-violation side. We emphasize here that each run
can involve an arbitrarily large number $M$ of pairs. Each run
therefore qualifies as a large statistical ensemble for purposes
of the Bell test. A violation that is observed with finite
probability (for an essentially infinite number of pairs) is
therefore sufficient to preclude any LHV explanation of the
observed results. Since testing LHV theories is precisely the
purpose of performing the Bell test, this probabilistic violation
constitutes a successful violation of Bell's inequality.

The above argument regarding the interpretation of the
probabilistic violation affects mainly the case $N=0$. The reason
is that averaging over a large number of runs results in a
violation for any finite $N$, leaving no caveats about the
observed violation. The same reason, however, makes the case $N=0$
of special importance for the present discussion of interpreting
the probabilistic violation. Even though averaging over many runs
shows no violation of Bell's inequality, the probabilistic
violation is sufficient to preclude LHV theories and therefore
constitutes a successful violation of Bell's inequality, as
mentioned above. It is also worth emphasizing here that when $N=0$
no nonlocal ancillary resource is used at all; any correlations
between the two observers are carried solely by the flying
particles.

One might wonder whether the finite violation probability observed
in our numerical calculations for $N=0$ is a result of the fact
that the number of pairs $M$ was finite (under 1000 in all of our
numerical calculations). It is straightforward to verify, however,
that this is not the case. As mentioned above, every time a pair
is measured and the results of both measurements are known, the
state of the ancillary resource evolves according to these
measurement outcomes. If one starts with $N=0$ and takes the
experimental runs where the first pair to be measured in the
$a'$-$b'$ bases gives $\sigma_{a'} \sigma_{b'}=+1$, one finds that
the ancillary resource evolves into a state equivalent to
$(\ket{10}+\ket{01})/\sqrt{2}$ (the only possible difference from
this state is the existence of some additional particles whose
location is known with certainty, which would happen when the
first few measurements are performed in bases other than
$a'$-$b'$). With this new initial state of the ancillary resource,
one finds that the ensemble average of all subsequently prepared
pairs will have $\gamma=1/2$ (i.e. on the violation side of the
inequality). Taking into consideration the fact that the range of
$C_{a',b'}$ is finite (from $-1$ to $1$), one can see that the
only way for the average over all the runs in this finite
sub-ensemble to be on the violation side of the inequality is to
have a finite fraction of all these runs being on the violation
side. Thus the finite violation probability for an infinite number
of runs is proved.

So far we have discussed a sequence of measurements using the same
ancillary resource. We now express explicitly how the above
analysis can be cast in terms of multipartite-entangled states. If
a large number $M$ of target-particle pairs are prepared before
performing any measurement, the state of the entire system would
be given by
\begin{widetext}
\begin{equation}
\ket{\Psi} = \sum_{n_1=0}^{1} \cdots \sum_{n_M=0}^{1}
\sum_{j=0}^{N} \sqrt{\frac{P_j}{2^M}} \ket{j+\sum_{k=1}^M
n_k,N-j+M-\sum_{k=1}^M n_k}_{\rm anc} \otimes \ket{n_1,1-n_1}
\otimes \cdots \otimes \ket{n_M,1-n_M},
\label{eq:Multipartite_state}
\end{equation}
\end{widetext}
where, for notational simplicity, the target-particle states
$\ket{g}$ and $\ket{e}$ are now expressed as $\ket{0}$ and
$\ket{1}$, respectively. For purposes of analyzing the outcomes of
measurements performed on the target particles, the
flying-particle and ancillary degrees of freedom can be traced
out, which still results in a multipartite entangled state for the
target particles. As the measurements proceed on target-particle
pairs, the state of the remaining pairs evolves according to the
initial measurement outcomes, resulting in the observed
probabilistic violation. In other words, the entanglement within
the unmeasured pairs increases or decreases, depending on the
measurement outcomes for the measured pairs.

\section{Conclusion}

In conclusion, we have considered the question of performing
Bell-type experiments using pairs that are entangled with each
other. We have presented a multipartite entangled physical system
where a violation of Bell's inequality would be obtained
probabilistically, with the violation or lack thereof being
decided by a sequence of quantum (random) outcomes. This
probabilistic violation is sufficient to preclude local-realistic
models.

%\begin{acknowledgments}
This work was supported in part by the National Security Agency
(NSA), the Army Research Office (ARO), the Laboratory for Physical
Sciences (LPS), the National Science Foundation (NSF) grant
No.~0726909, the Austrian Science Foundation FWF within Project
No. P19570-N16, SFB and CoQuS No. W1210-N16 and the European
Commission, Project QAP (No. 015848).
%\end{acknowledgments}


\begin{thebibliography}{99}
\bibitem{JustifyBases} The choice of bases used here gives a maximum
violation value of $5/2$, which is somewhat smaller than the
widely known value of $2\sqrt{2}$. However, in order to simplify
the task of keeping track of the multipartite quantum
correlations, it is desirable for our analysis to use the
$\left\{\ket{0},\ket{1}\right\}$ basis (for both observers) as one
of the possible measurement bases. With this choice, three out of
four possible measurement settings will contain a
$\left\{\ket{0},\ket{1}\right\}$ basis. The resulting
simplification will become clear in our analysis.
\bibitem{EntanglementArgument} The fact that the BEC can be reused
indefinitely was used in Ref.~\cite{Ashhab1} to argue that the
Bell-inequality violation truly probes the flying-particle
entanglement, as oppposed to the entanglement between the two
parts of the ancillary BEC. An alternative argument that
demonstrates this fact arises when the two ancillary condensates
are small parts extracted from a much larger BEC. The entanglement
between the two reference condensates in this case is inversely
proportional to the size of the large BEC. By taking the limit of
an infinitely large BEC, one can have two good phase-reference
states that are not entangled with each other.
\end{thebibliography}
\end{document}